\documentclass[11pt,letterpaper]{article}
\pdfoutput=1
\usepackage{jcappub}
\usepackage{amsmath}
\usepackage{graphicx}
\usepackage[config=altsf]{subfig}
\usepackage{feynmf}
\usepackage{array}
\usepackage{slashed}
\usepackage{afterpage}
\usepackage{appendix}
\usepackage{multirow}
\usepackage{nicefrac}

\newcommand{\be}{\begin{equation}}
\newcommand{\ee}{\end{equation}}
\newcommand{\ba}{\begin{eqnarray}}
\newcommand{\ea}{\end{eqnarray}}

\definecolor{darkred}{RGB}{175,0,0}
\definecolor{darkblue}{RGB}{0,0,175}

\newcommand{\citeref}[1]{Ref.~\cite{#1}}
\newcommand{\citerefs}[1]{Refs.~\cite{#1}}

\usepackage{color}

\graphicspath{{./figs/}}

\title{Make Dark Matter Charged Again}
\author{Prateek Agrawal,}
\author{Francis-Yan Cyr-Racine,}
\author{Lisa Randall}
\author{and \\Jakub Scholtz}
\affiliation{Department of Physics, Harvard University, \\ Cambridge, MA 02138, U.S.A.}

\emailAdd{prateekagrawal@fas.harvard.edu, fcyrraci@physics.harvard.edu,  randall@physics.harvard.edu, jscholtz@physics.harvard.edu}

\abstract { We revisit constraints on dark matter that is charged
  under a $U(1)$ gauge group in the dark sector, 
decoupled from Standard Model forces.  We find that the strongest
constraints in
  the literature are subject to a number of mitigating factors.
  For instance,  the naive dark matter thermalization timescale in
  halos is corrected by saturation effects that slow down
  isotropization for modest ellipticities.  The weakened bounds
  uncover interesting parameter space, making models with weak-scale
  charged dark matter viable, even with electromagnetic strength
  interaction. This also leads to the intriguing
  possibility that dark matter self-interactions within small dwarf
  galaxies are extremely large, a relatively unexplored regime in
  current simulations. Such strong interactions suppress heat transfer
  over scales larger than the dark matter mean free path, inducing a
  dynamical cutoff length scale above which the system appears to have
  only feeble interactions. These effects must be taken into account
  to assess the viability of darkly-charged dark matter. 
 Future analyses and measurements should probe
  a promising region of parameter space for this model.
}

\arxivnumber{1610.04611}

\begin{document}
\maketitle

\section{Introduction}

Apart from its manifold gravitational influences,  dark  matter has so
far eluded detection, prompting model builders to think more broadly
about what dark matter can be and in the process consider other and
more subtle ways to search for it.  One intriguing possibility has
always been charged dark matter, with dark matter charged under its
own force
that Standard Model particles do not necessarily experience. This can
constitute all of the dark matter, as has been considered in 
\citerefs{Goldberg:1986nk,HOLDOM198665,1992ApJ...398..407G,
1992ApJ...398...43C,Foot:2004pa,Feng:2008mu,Ackerman:2008gi,
Feng:2009mn,Feng:2009hw,ArkaniHamed:2008qn,Kaplan:2009de,2010PhRvD..81h3522B,
Kaplan:2011yj,Behbahani:2010xa,Das:2012aa,Hooper:2012cw,
Aarssen:2012fx,Cline:2012is,Tulin:2013teo,Tulin:2012wi,
Baldi:2012ua,Cyr-Racine:2013ab,Cline:2013zca,Chu:2014lja,
Cline:2013pca,Bringmann:2013vra,Gabrielli:2013jka,Boddy:2014yra,Archidiacono:2014nda,
Choquette:2015mca,Buen-Abad:2015ova,Ko:2016fcd,Bringmann:2016ilk}, 
or it
can be a fraction of the dark matter, as was studied in 
\citerefs{Fan:2013tia,Fan:2013yva,
Cyr-Racine:2013fsa,Chacko:2016kgg}.  More generally,  research 
on self-interacting dark matter  has been motivated in part by
potential discrepancies in conventional dark matter scenarios,
especially  on small scales. However, most papers on charged dark
matter imply that it is very restricted by current constraints.

It is of interest to ask whether dark matter is necessarily
constrained to have only very tiny or short-ranged interactions. An
immediate concern is that new interactions in the dark sector --
especially those involving massless particles -- could significantly
delay the epoch of dark matter kinetic decoupling and affect the
large-scale structure of the Universe. However, for dark matter with a
weak-scale mass, such bounds are generally much weaker
\cite{Kesden:2006vz,Ackerman:2008gi,Feng:2009mn,Schneider:2016uqi}
than those coming from collapsed dark matter structures such as
clusters, galaxies and dwarves. This paper thus focuses on these
latter objects.

Charged dark matter -- even when charged under an invisible new
$U(1)$ -- can potentially lead to observable differences in the dark
matter density distribution that can be helpful when searching for
interactions (see e.g. \cite{Vogelsberger:2012ku,Zavala:2012us,Vogelsberger:2012sa,Vogelsberger:2014pda,Boehm:2014vja,Buckley:2014hja,Kaplinghat:2015aga,Schewtschenko:2014fca,2016MNRAS.460.1399V}), but can also in principle impose serious constraints. This model, which we shall refer to here as darkly-charged dark matter, is simple from a particle physics perspective but leads to extremely interesting consequences in astrophysics. Constraints arise from the triaxiality of galaxy and cluster-scale dark matter halos, since overly strong interactions can wipe out deviations from isotropy. A somewhat weaker constraint can be derived from observations of merging galaxy clusters \cite{Markevitch:2003at,Randall:2007ph,Dawson:2012fx,Mohammed:2014iya,Dawson:2014jca,Kahlhoefer:2013dca,Kahlhoefer:2015vua,Harvey:2015hha,Massey:2015dkw,Randall:2016nxa,Robertson:2016xjh}. A further constraint comes from the survival of
dwarf galaxies as they move through the halo \cite{Kahlhoefer:2013dca,Kahlhoefer:2015vua,2016MNRAS.461..710D}.  

In this paper we revisit the constraints on dark matter charged under
its own $U(1)$ to show that the allowed parameter space is broader
than implied by the literature. We show explicitly that even with
pre-existing assumptions, the literature overestimates the bound by
about an order of magnitude.  Furthermore these constraints might be
less reliable than assumed due to uncertainties in initial conditions,
for example. Our result is important because it opens the window for
charged dark matter as light as the weak scale, even if it carries a
charge as strong as that of electromagnetism.  It is also of interest
for Double-Disk dark matter~\citeref{Agrawal:2017rvu} as it opens the
possibility of all dark matter being charged while allowing a dark
matter disk. We also suggest interesting implications for charged dark
matter that have not yet been investigated, including using the radial
dependence of ellipticity to disentangle the impact of velocity
anisotropies from that of anisotropic potentials, and the
modification of the dwarf galaxy populations in both mass and internal
structure, with important consequences for small-scale issues
\cite{2010AdAst2010E...5D,Gentile:2004tb,BoylanKolchin:2011de,Walker:2011zu}.

Dark matter self-interactions through the dark photon are strongly
enhanced for low velocity systems such as dwarf galaxies. The coupling strengths allowed by our revised constraints lead to extremely large cross sections in dwarf galaxies -- a relatively
unexplored regime in self-interaction dark matter (SIDM) simulations -- while being roughly
consistent with the desired SIDM cross sections required to solve potential issues at galaxy and cluster scales \cite{Kaplinghat:2015aga}. Generally, many SIDM models invoke a massive mediator to cut off strong
interactions at low velocities. 
Interestingly, our very strong interactions can inhibit heat
transfer over scales larger than the dark
matter mean free path, inducing a dynamical cutoff length scale above
which the system appears to have only feeble interactions
\cite{Gnedin:2000ea,Moore:2000fp,2003JKAS...36...89A,Ahn:2004xt}. 
Such a dynamical cutoff would likely obviate the need of introducing
the cutoff through the mass of the mediator.

The outline of this paper is as follows. We introduce the model in
section~\ref{sec:model}. In section~\ref{sec:ellipticity} we study the
strongest constraint quoted in the literature, the ellipticity
constraint, and point out reasons why the actual bounds might be
weaker. This makes the constraints coming from dwarf galaxies the most
stringent, which we study in section~\ref{sec:dwarf-survival}. We briefly discuss the constraints from merging galaxy clusters in section~\ref{sec:bullet-cluster}. We highlight future work in section~\ref{sec:future}, emphasizing the importance of the strongly interacting regime at low velocities in section \ref{sec:superstrong}, and finally conclude in
section~\ref{sec:conclusions}. 

\section{Model}
\label{sec:model}
We consider a simple model with a Dirac fermion $X$ of
weak-scale mass $m_X$, charged under a new dark $U(1)$ gauge symmetry with gauge boson $\gamma_D$.
The
Lagrangian is
\begin{align}
  \mathcal{L}
  &=
  -\frac14 V_{\mu\nu} V^{\mu\nu} + \bar{X} i \slashed{D} X - m_X \bar{X} {X}, 
\end{align}
where $V_\mu$ is the dark photon and $D_\mu = \partial_\mu + i g_D
V_\mu$. We also define $4\pi \alpha_D = g_D^2$. 
We assume that there is no kinetic mixing between photons and the dark
photons. In order to avoid conflict with cosmic microwave background (CMB) measurements of the abundance of relativistic species  \cite{Ade:2015xua}, we take the dark photon bath temperature $T_D$ to be half that of the CMB photon, $\xi \equiv T_D/T_{\rm SM} = 0.5$. In this simple model, the dark matter relic abundance can be
set by a thermal freezeout, which we briefly review in the next section.

\subsection{Relic abundance}\label{sec:relic_abund}

The relic abundance is determined by the freezeout process $X\bar{X}
\to \gamma_D\gamma_D$ with a leading thermally averaged cross-section given by
\begin{align}
  \langle \sigma v_\text{M\o l} \rangle 
  &= 
  \frac{\pi \alpha_D^2}{m_X^2} \bar{S}_{\rm ann}(\alpha),
\end{align}
where we follow \citeref{Gondolo:1990dk} definition of M\o ller
velocity. The Sommerfeld enhancement $\bar{S}_{\rm ann}$ is important for
$\pi \alpha_D / v_f \gtrsim 1$, which arises 
for large $m_X \gtrsim 1$ TeV.
We use the results of \citeref{vonHarling:2014kha} to
include the thermally averaged Sommerfeld enhancement.

\citeref{Gondolo:1990dk} gives the corresponding relic density for
this cross-section:
\begin{align}
\Omega_X =
    \frac{16\pi^3}{9\sqrt{5\pi}}
    \frac{g_0^*}{\sqrt{g_\text{eff}}}
    \frac{T_0^3}{M_\text{pl}^3}
    \frac{x_f}{\langle \sigma v_\text{M\o l}\rangle H_0^2},
\end{align}
where $T_0$ is the CMB temperature today, $H_0$ the Hubble constant
today, $M_\text{pl}$ is the Planck mass, $g_0^*$ is the effective number of relativistic degrees of
freedom today, while $g_\text{eff}$ is the effective number of degrees
of freedom at freezeout, which happens when $T_f = m_X/x_f$. We can
determine $x_f$ as a solution to the equation
\begin{equation} 
  \xi^{3/2} \frac{\sqrt{45}}{8\pi^2}
  \frac{g_X}{\sqrt{g_{*}(T_f)}} \frac{\alpha_D^2 M_\text{pl}}{m_X}
  \delta(\delta+2) = \omega=\sqrt{x_f} e^{x_f/\xi}, 
\end{equation} 
where we denote the number of degrees of freedom in $X$ by $g_X$ and
$\delta \sim 1.5$ is a matching parameter to the exact numerical
solution ($x_f$ is only logarithmically sensitive to $\delta$). We use
the approximate solution~\cite{Feng:2008mu},
\begin{equation}
x_f = \xi \log \omega - \frac{1}{2}\xi \log (\xi \log \omega),
\end{equation}
and require that $\Omega_X = 0.265$ \cite{Ade:2015xua} in
order to obtain $\alpha_D$ as a function of $m_X$. We note that
it is both possible to decrease and increase $\alpha_D$ for a given
value of $m_X$. We can
decrease $\alpha_D$ by opening another annihilation channel for
$X$, and increase $\alpha_D$ by introducing additional
contributions to $\Omega_\text{DM}$.
Assuming no other freezeout channels and that $X$ is the only
contribution to $\Omega_{\text{DM}}$, we show our result for relic
abundance in 
figure~\ref{fig:constraints}.

\section{Constraints and Phenomenology}

\subsection{Ellipticity}
\label{sec:ellipticity}

One of the strongest purported constraints on charged dark
matter comes from observed triaxial structure of galaxy halos.
Self-interacting dark
matter can in principle reduce the degree of anisotropy by creating a
more isotropic dark matter velocity distribution. By measuring
non-zero ellipticity of the gravitational potential of NGC720, 
\citeref{Buote:2002wd} have made it possible to obtain constrains on
self-interaction
strength of dark matter. \citeref{Feng:2009mn} have
used this data to put bounds on self-interacting dark matter. These
authors considered both hard interactions, where large momentum
exchange immediately reduces anisotropy significantly and soft
exchanges, where the cumulative effect of many interactions of a dark
matter particle as it passes through the galaxy serves to make the
velocity distribution more isotropic. The latter is the more
important process: even though each collision changes the momentum of
a particle by
 a small amount a single dark matter particle can scatter many
times during its orbits over the lifetime of the galaxy. 

We review the calculation in the literature, highlighting aspects
which modify the bounds.
The usual calculation involves obtaining the characteristic
timescale for an average particle to change its kinetic energy by an
$\mathcal{O}(1)$ factor, and interpreting this as the 
timescale $\tau_{\rm iso}$ on which the velocity vector fully randomizes.
We present the detailed calculation below; the final result is
\cite{Feng:2009mn} 
\begin{equation}
  \tau_{\rm iso}
  = 
  \mathcal{N}
  \frac{m_X^3 v_0^3}{\alpha_D^2 \rho_X} 
  \left( \log \frac{(b_{\text{max}} m_X v_0^2/\alpha_D)^2+1}{2}\right)^{-1} 
  =  
  \mathcal{N}\,
  \frac{m_X^3 v_0^3}{\alpha_D^2 \rho_X} 
  (\log \Lambda)^{-1}.
  \label{eq:taufeng}
\end{equation}
We comment on the numerical prefactor $\mathcal{N}$ below. There are
two qualitatively important aspects of our analysis:
\begin{itemize}
  \item In NGC720 the baryonic component dominates the enclosed gravitational
    mass until about $r \sim 6$ kpc~\cite{2011ApJ...729...53H}.  Therefore, ellipticity
    measurements within this radius do not constrain
    the dark matter potential. We choose to apply the ellipticity
    constraint at $r=6$ kpc where the isotropization rate is
    relatively smaller due to a lower dark matter density. 

  \item The long range interactions between dark matter are 
    screened at the inter-particle
    spacing in the galaxy (or galaxy cluster). 
    Typically the screening  length is taken to be the Debye screening length $\lambda_D$. However, in a neutral plasma with equal mass opposite charges, if the inter-particle spacing $\lambda_{pp} \ll \lambda_D$, then the contributions to scattering from individual charges cancel\footnote{One can alternatively see this as scattering on dipoles, which do not contribute to the logarithmic (dominant) part of the cross-section integral.} leaving behind terms of order $\lambda_{pp}/\lambda_D$. As a result it is appropriate to take the inter-particle spacing as the screening length

\end{itemize}

Our numerical prefactor is 
\begin{align}
  \mathcal{N} 
  = \frac{3 }{16\sqrt{\pi} }.
\end{align}
While we agree on the functional form of the timescale, this prefactor
is somewhat different from \citeref{Feng:2009mn}. 
This difference is accounted for by a different choice of $r$ (3
kpc), IR cutoff (Debye mass) and some other numerical
factors (normalization of velocity, cross section and energy transfer)
in that work. 
Consequently, our isotropization timescale is larger by a factor
\begin{equation}
  \underbrace{\frac{3}{2}}_{\log \Lambda} 
  \times \underbrace{\frac{3}{1}}_{\rho} 
  \times  \underbrace{\frac{1}{4}}_{d\sigma/d\Omega} 
  \times \underbrace{\frac{3}{2}}_{\langle v^2 \rangle 
  = 3v_0^2/2} \times \underbrace{\frac{2}{1}}_{\delta E_k}
  = \frac{27}{8} 
  \; .
  \label{eq:factors}
\end{equation}
Moreover, there are additional considerations whose numerical effects
require explicit additional calculations we will present in
section~\ref{sec:diffeq-velani}~and~\ref{sec:ellipticityradius}:
\begin{itemize}
  
  \item It is not sufficient to simply calculate the interaction rate,
    or even the rate at which energy transfers from one velocity
    component to another. The rate at which the interaction occurs is
    sensitive to velocity anisotropy.
    As the initially smaller component of the velocity grows comparable to the larger one, the rate of energy
    transfer slows down. (Otherwise the smaller one would continue to
    grow indeterminately which of course is not the case.) This saturation effect can relax the bounds from 
    ellipticity significantly.

  \item Furthermore, the constraint depends on the radius at which the
    ellipticity is measured. This is important because the best
    ellipticity measurements apply in the outer regions of galaxies
    where the density is lowest and therefore interactions are less
    frequent. 

\end{itemize}
Finally there are additional challenges to ellipticity measurements as constraints on dark matter self-interactions that shed doubt on the viability of ellipticity measurements as constraints on dark matter self-interaction rate:
\begin{itemize}

  \item  The measured ellipticity of the gravitational potential is due to anisotropy in the density of the
   dark matter distribution. All authors constraining loss of ellipticity calculate the rate of energy transfer, which might apply to heat flow among different radii or between different velocity components. We can estimate the rate at which  the velocity distribution becomes isotropic through
   a calculation, but the rate at which the density distribution (and the gravitational potential) becomes isotropic
   requires N-body simulations, as discussed in Ref.~\cite{Peter:2012jh}. In isolated static haloes one can argue that the approach to isotropy is the same for velocity distribution and density distribution because the same 
   collisional term contributes to the change of the distribution function. However, substructure, dark matter
   streams, accretion, rotation or long relaxation time can change this expectation. As a result, the ellipticity
   bound on dark matter self-interaction may be weaker than that obtained from the requirement on isotropy
   in velocity distribution alone.
  \item The constraint from galaxies is stronger than the constraint
    from galaxy clusters. But the former relies on only a single
    galaxy  NGC720, which might not be representative. As
    this galaxy does not exhibit strong ellipticity, it might be
    undergoing isotropization at a diminished rate as the velocity
    components become more comparable. In any case, having access to a larger statistical sample of galaxy ellipticities is key to make this type of constraint robust. 
 
  \item The history of the galaxy is a key input that observations do
    not have access to. Recent mergers can contribute to a large
    ellipticity even in the presence of strong interactions. Even if a
    deviation from the cold dark matter (CDM) prediction is expected, ellipticity observations
    are sensitive to the initial distribution, so robust conclusions
    can only be drawn from observables that distinguish CDM and Darkly Charged dark matter
    independently of initial conditions.
\end{itemize}

Given the latter items, even  our more careful analysis very likely
overstates the constraint as we  calculate the rate at which the
velocity components equilibrate but even a galaxy with isotropic
velocity distribution can exhibit ellipticity in its potential. Nonetheless, because it
might be useful for future analyses, and also to show that weak scale dark
matter is viable even if isotropy of velocities in galaxies is a serious constraint, we present the more careful computation here.

\subsubsection{Review of the timescale calculation}

In this section we review the calculation of the timescale for an
average dark matter particle to change its kinetic energy $E_k$ by an
$\mathcal{O}(1)$ factor. We define this timescale as
\begin{align}
  \tau_{\rm iso}
  &= 
  \frac{\langle E_k \rangle}{\langle \dot{E_k} \rangle},
\end{align}
where the $\langle \cdot \rangle$ denotes thermal averaging.
The average
rate of change of energy of a typical particle is
\begin{align}
  \langle\dot{E}\rangle 
  &= \int d\Omega\, dv \frac{d\sigma}{d\Omega} f(v)
  \,\delta E_k\, v \,n_X 
  \, ,
\end{align}
where $n_X$ is the number density of $X$ particles. For two equal mass particles with velocities $v_1$ and $v_2$, the energy loss/gain after a collision is $\delta E_k = \pm(1-\cos \theta_\text{cm})m(v_1^2-v_2^2)/4$, where $\theta_{\rm cm}$ is the final scattering angle in the center-of-mass (CM) frame.
The differential cross section is (see Appendix~\ref{sec:xsection})
\begin{equation}
  \frac{d\sigma}{d\Omega}  
  = \frac{4\alpha_D^2}{m_X^2 |v_1-v_2|^4 (1-\cos \theta_{\text{cm}})^2}.
\end{equation}
We assume a Maxwell-Boltzmann distribution with dispersion $v_0$ for
the dark matter velocities. This yields
\begin{align}
  \langle\dot{E}\rangle 
  &=
  \frac{4\pi\alpha_D^2 \rho_X  }{2m_X^2} 
  \int_{\theta_\text{min}}
  \frac{1-\cos \theta_\text{cm}}{(1-\cos\theta_\text{cm})^2} d\cos\theta_\text{cm}  \int \frac{4 v
  dv}{\sqrt{\pi} v_0^3} \exp\left(-\frac{v^2}{v_0^2}\right)\,,
\end{align}
where $\rho_X = m_Xn_X$ is the dark matter energy density. Since the leading contribution from positively and negatively
charged particles will cancel, we use the inter-particle spacing
($\lambda_P = (m_X/\rho_X)^{1/3}$) as an infrared cutoff for the impact parameter.
The
relationship between largest impact parameter $b_{\rm max}=\lambda_P$ and angle of scattering is
\begin{align}
  \frac{\theta_{\rm min}}{2} 
  &= \cot^{-1} \frac{ \lambda_P m_X v_0^2}{\alpha_D}.
\end{align}
Therefore, 
\begin{align}
  \langle\dot{E}\rangle 
  &=\frac{4\sqrt{\pi}}{ v_0}  \frac{\alpha_D^2 \rho_X }{m_X^2}
  \log \left[1+ \frac{ \lambda_P^2 m_X^2 v_0^4}{\alpha_D^2}\right] 
  \equiv\frac{4\sqrt{\pi}}{ v_0}  \frac{\alpha_D^2 \rho_X }{m_X^2}
  \log \Lambda\,.
\label{eq:lambda}
\end{align}
The timescale to make the velocity distribution isotropic is then
\begin{align}
  \tau_\text{iso}
  &= 
  \frac{3}{16\sqrt{\pi}}
  \frac{ m_X^3 v_0^3}{\alpha_D^2 \, \rho_X\log \Lambda  }
  \, .
  \label{eq:tauc}
\end{align}
As noted above, the prefactor differs from the calculation
in~\cite{Feng:2009mn} by the factor outlined in
Eq.~(\ref{eq:factors}).  

This calculation gives the rate at which a large initial ellipticity
decreases, but does not
suffice to determine the time it takes to erase relatively small
ellipticities, for which all velocity components are substantial and
comparable. We address this issue in the next section, where we
explicitly calculate the time to erase ellipticity for different
measured ellipticity values.

\subsubsection{Evolution of the velocity anisotropy}
\label{sec:diffeq-velani}

In this section, we illustrate how the transfer of energy from a hot population to a colder
one evolves as a function
of time. This would apply either to energy transferring over different radii or for
energy transfer from a velocity distribution in one direction to a velocity component along another axis. Here we model ellipticity (defined by the ratio of lengths of
semi-minor axis, $b$, and semi-major axis, $a$) as an anisotropy in
velocity distribution (a strong assumption). In particular, in a
virialized halo, $\langle v^2 \rangle \sim R^{-1}$ and therefore:
\begin{align}
  \epsilon 
  &= 
  1 - \frac{b}{a} \sim 1 - \frac{\langle v_a^2\rangle}{\langle v_b^2\rangle}
  \, .
\end{align}
The timescale calculation above estimates the growth of ellipticity
only when $\langle v_a^2 \rangle \ll \langle v_b^2 \rangle$,
that is for large ellipticities $\epsilon \lesssim 1$. However, when
$\langle v_a^2 \rangle \lesssim \langle v_b^2 \rangle$, we expect a
much smaller growth of the subleading velocity component because the
process is proportional to the velocity anisotropy: 
\begin{equation}
  \frac{d \langle v_a^2 \rangle}{dt} 
  \propto 
  \left(\langle v_b^2 \rangle - \langle v_a^2 \rangle\right)^\gamma,
\end{equation}
where the index $\gamma$ depends on the type of scattering that is erasing anisotropy. Here, we use a slightly simpler physical system for the sake of clarity to
explore this slowing of the decline of ellipticity quantitatively. We
study a cold isotropic population of gas with a Maxwellian velocity
distribution, with a velocity dispersion $v_c$. This gas is placed in
a hotter bath, with the corresponding dispersion $v_h\gg v_c$. We study the
growth of the velocity dispersion of the colder population.

The Boltzmann equation reads
\begin{equation}
  \frac{d f(v_1)}{dt} = \int d^3 v_2 d\Omega' |v_1 - v_2| \frac{d\sigma}{d\Omega'} \left( f(v_1)f(v_2) - f(v_3)f(v_4)  \right).
  \label{eq:boltzmann1}
\end{equation}
In the above equation the final velocities $v_3$ and $v_4$ are uniquely determined by $v_1$, $v_2$ and the scattering angle $\Omega'$. The unprimed quantities are in the galaxy frame, primed quantities are
in the CM frame. The definition of the probability distribution function $f(v)$ is
\begin{equation}
  f(v) = \frac{n_c}{\pi^{3/2} v_c^3} \exp\left(-\frac{v^2}{v_c^2}\right) + \frac{n_h}{\pi^{3/2} v_h^3} \exp\left(-\frac{v^2}{v_h^2}\right). 
\end{equation}
\begin{figure}[tp]
  \centering
  \includegraphics[width=0.70\textwidth]{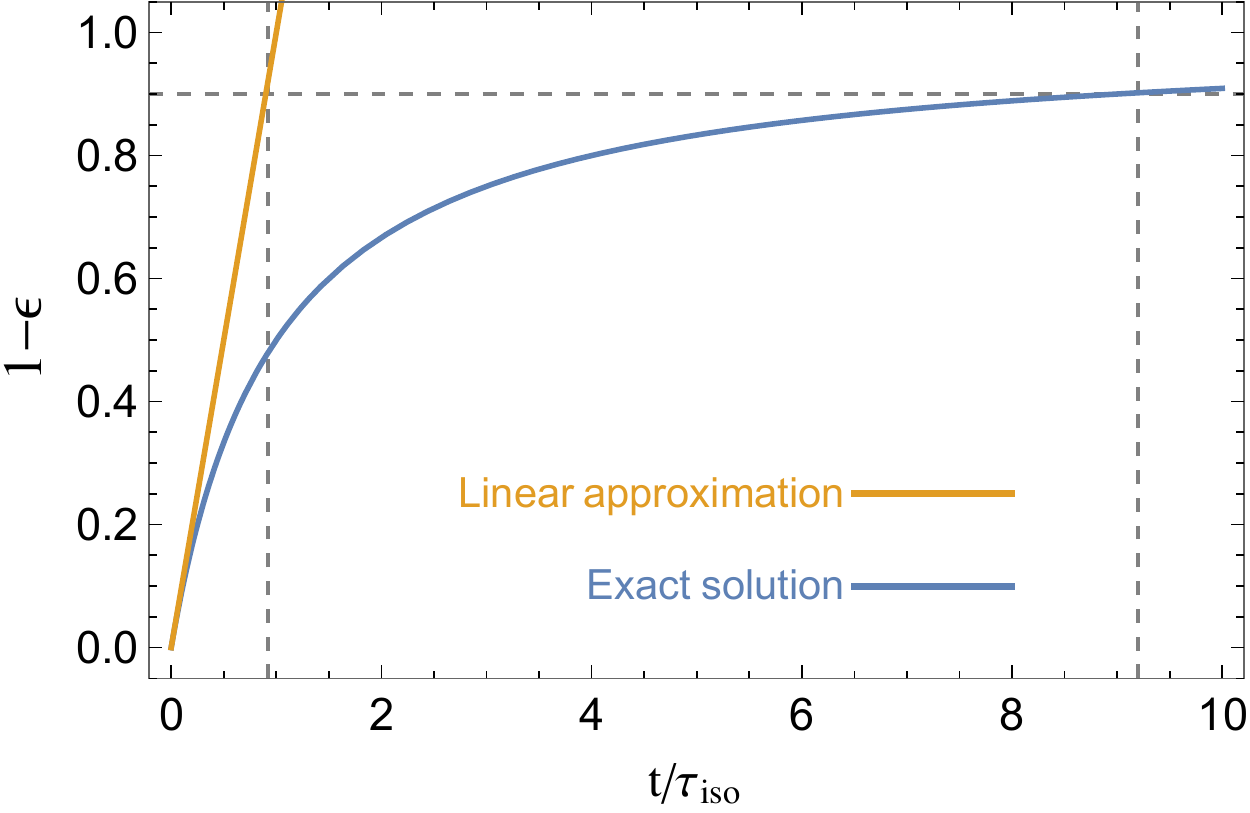}
  \caption{Time evolution of the ellipticity as modeled by our
    simplified system in section~\ref{sec:diffeq-velani} starting from an ellipticity of order unity. The solid orange line shows the approximate linear evolution, similar to the expansion performed in eq.~\eqref{eq:approx_vec_evol}. The solid blue
    curve illustrates the exact solution given in eq.~\eqref{eq:ellipt_evol_exact}. The dashed vertical lines illustrate the time to reach an ellipticity of 0.1 in both cases.  The saturation effect of the rate of isotropization as the halo becomes more isotropic is clearly visible in the exact solution. 
  }
  \label{fig:ellvst}
\end{figure}
Since the distribution is uniquely determined by $v_c$, it is
sufficient to look at the evolution of the $v_1 = 0$ bin. Since we are setting $v_1 = 0$, the kinematics simplifies and the final
velocities in the galaxy frame are
\begin{align}
  v_{3,4}^2 
  &= \frac{1}{2}v_2^2
  (1\pm\cos \theta').
\end{align}
Finally, the differential cross-section is again
\begin{equation}
  \frac{d \sigma}{d\Omega} = \frac{4\alpha_D^2}{m_X^2 |v_1-v_2|^4 (1-\cos\theta')^2} .
\end{equation}
With all the pieces ready, we can focus on the evolution of the $v_1 = 0$ bin. The Boltzmann eq.~(\ref{eq:boltzmann1}) tells us (using
$\cos \theta' = x$):
\begin{align}
  \!\!-\frac{3 \dot{v}_c n_c}{\pi^{3/2} v_c^4}
  -\frac{3 \dot{v}_h n_h}{\pi^{3/2} v_h^4}
  &= 
  \frac{4\alpha_D^2}{m_X^2  }
  \int \frac{ d^3 v_2}{ v_2^3} 
  \frac{d\phi' dx }{(1-x)^2}
  \left(f(0) f(v_2) - f(v_2\sqrt{(1-x)/2}) f(v_2\sqrt{(1+x)/2})\right).
  \label{eq:boltzmann2-mod}
\end{align}
We will neglect the backreaction on the bath, $\dot{v}_h$.
Using rotation invariance, we can perform the angular $v_2$ integral
and the $\phi'$ integral. Hence,
\begin{align}
  -\frac{3 \dot{v}_c}{ v_c^4}
  = &
  \frac{32 \pi^3\sqrt{\pi}\alpha_D^2}{m_X^2 n_c }
  \int    \frac{ d v_2}{ v_2} \frac{dx}{(1-x)^2}
  \left(f(0) f(v_2) - f(v_2\sqrt{(1-x)/2}) f(v_2\sqrt{(1+x)/2})\right).
  \label{eq:boltzmann2b-mod}
\end{align}
Rather than do the exact integral, we focus instead on the $\theta \to
0$ ($x \to 1$) limit, which dominates the behavior. After some algebra
(see appendix~\ref{sec:odeapp}) the Boltzmann equation reads
\begin{align}
  \dot{v}_c
  &= -\frac{8\sqrt{\pi}\alpha_D^2 n_h}{3m_X^2 v_h^5 v_c }  
  \left( v_c^2 - v_h^2\right)^2 \int_0^{\cos^{-1}\theta_\text{min}} 
  \frac{dx}{(1-x)}.
\end{align}
As before, we use the inter-particle
spacing to cut off the infrared divergence above. The result
is a first order differential equation for $v_c$
\begin{equation}
  v_c \dot{v}_c = -\frac{8\sqrt{\pi}\alpha_D^2 n_h}{3m_X^2 v_h^5 }  \left( v_c^2 - v_h^2\right)^2 \log \Lambda,
\end{equation}
where $\Lambda$ is the same as in eq.~(\ref{eq:lambda}), with $v_0$ replaced by $v_h$. The solution takes the form
\begin{equation}
  v_c^2(t) = v_h^2-\frac{v_h^2}{\frac{t}{\tau}+\frac{v_h^2}{v_h^2 - v_{c,0}^2}}
  \,,
\end{equation}
where $v_{c,0}$ is the initial velocity dispersion of the cold population, and where we have defined an effective timescale to isotropize the velocity distribution:
\begin{align}
  \tau =  \frac{3 m_X^2 v_h^3}{16\sqrt{\pi}\alpha_D^2 n_h \log \Lambda}  
  \label{eq:tau}
  \, . 
\end{align}
The ellipticity evolves over time as
\begin{equation}\label{eq:ellipt_evol_exact}
\epsilon(t) \sim 1 - \frac{v_c^2}{v_h^2} = \frac{\epsilon_0}{\frac{t}{\tau}\epsilon_0+1}.
\end{equation}
We can see that for small initial dispersion $v_{c,0} \ll v_h$ and  $t
\ll \tau$, 
\begin{equation}\label{eq:approx_vec_evol} v_c^2 \sim  v_h^2 \frac{t}{\tau} =
  \frac{16\sqrt{\pi}\alpha_D^2 n_h \log\Lambda}{3 m_X^2 v_h} \times t.
\end{equation}
However, this early time expansion neglects the saturation effect as
the temperature of the cooler component gets comparable to the hotter one.
In figure~\ref{fig:ellvst} we show that the time taken to get to a
small ellipticity can be much longer than the timescale in
eq.~\ref{eq:tau}.
In order to include this effect we need to determine the degree to
which the velocity components become isotropic inside the galaxy.
Since ellipticity in NGC720 varies with the galactic radius we need to
extend our analysis to include radius dependence, which we now turn to
in the next section.

\begin{figure}[t]
  \begin{center}
    \includegraphics[width=0.8\textwidth]{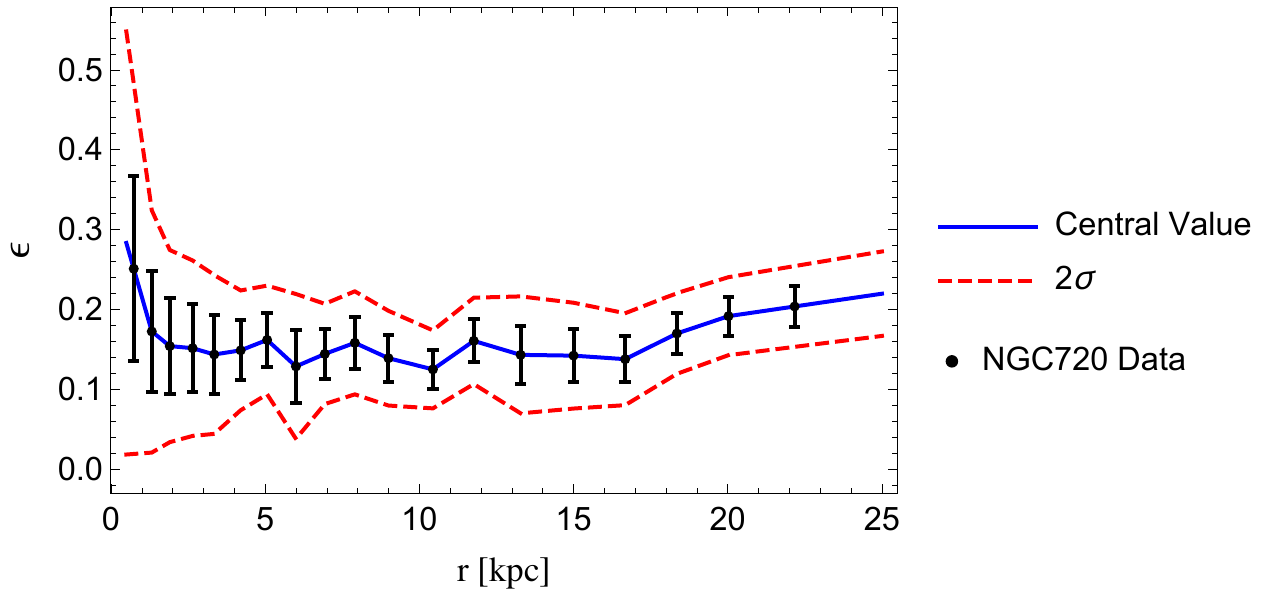}
  \end{center}
  \caption{\label{fig:NGCellipticity} Ellipticity of the NGC720
  potential as measured by Ref.~\cite{Buote:2002wd}. The black data points show
  the measurements with $1\sigma$ error bars. The
blue curve is our interpolation of their central values, while the $2\sigma$ error bands are in dashed red.}
\end{figure}

\subsubsection{Ellipticity and Density as a function of radius}
\label{sec:ellipticityradius}

Measuring ellipticity as a function of radius is a complex process. To
tackle this challenge, authors of \citeref{Buote:2002wd} have adopted
an iterative process based on measuring second moments inside
elliptical annuli. We show their measurements for NGC720 in
figure~\ref{fig:NGCellipticity}. 
From eq.~(\ref{eq:tauc}) we see that regions with the highest phase
space density $\rho/v_0^3$ yield the shortest time scales.  We use the
data of \citeref{2011ApJ...729...53H} to determine both the dark
matter density and local virial velocity as a
function of radius. The shortest times and therefore the strongest
constraint would come the densest regions -- 
the inner most parts of the galaxy. However, the uncertainty on
ellipticity is significant for these smaller radii, weakening the
overall bound.

\begin{figure}
  \begin{center}
    \includegraphics[width=0.475\textwidth]{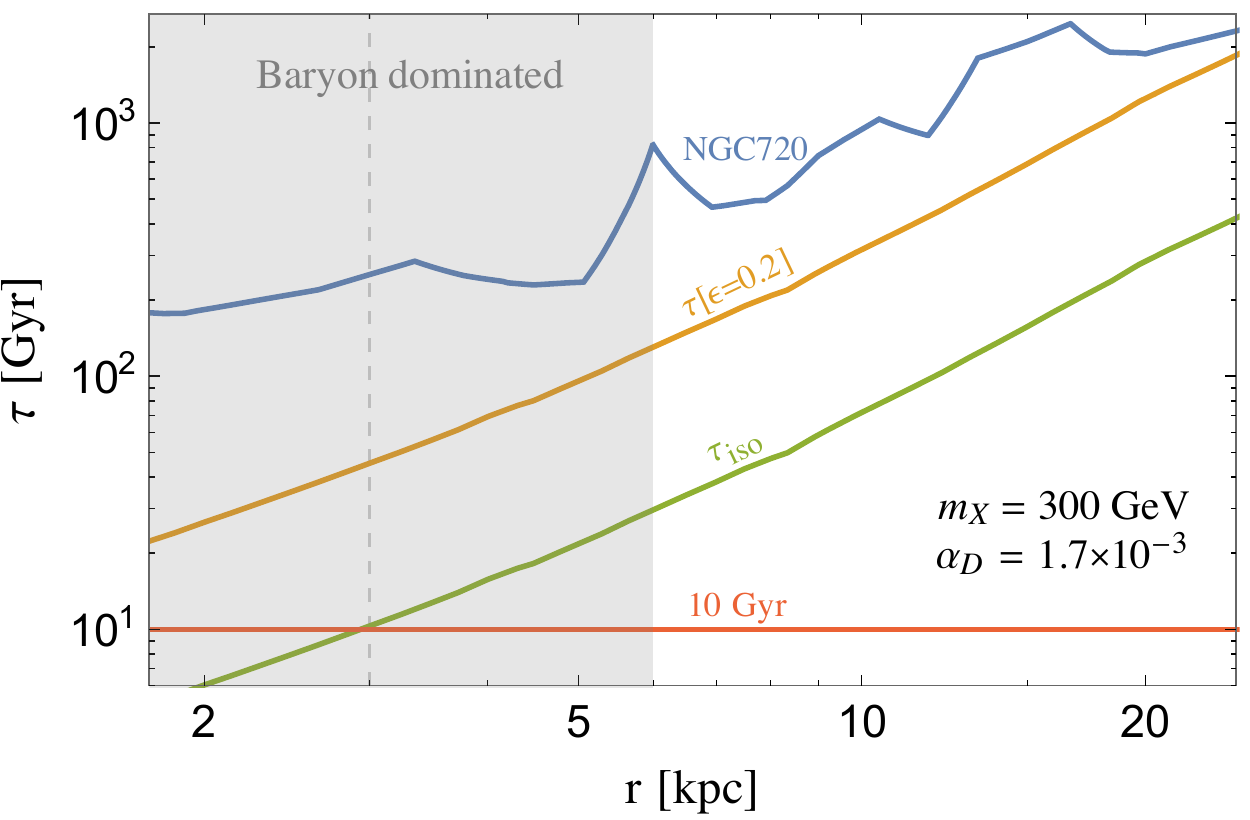}\qquad
    \includegraphics[width=0.475\textwidth]{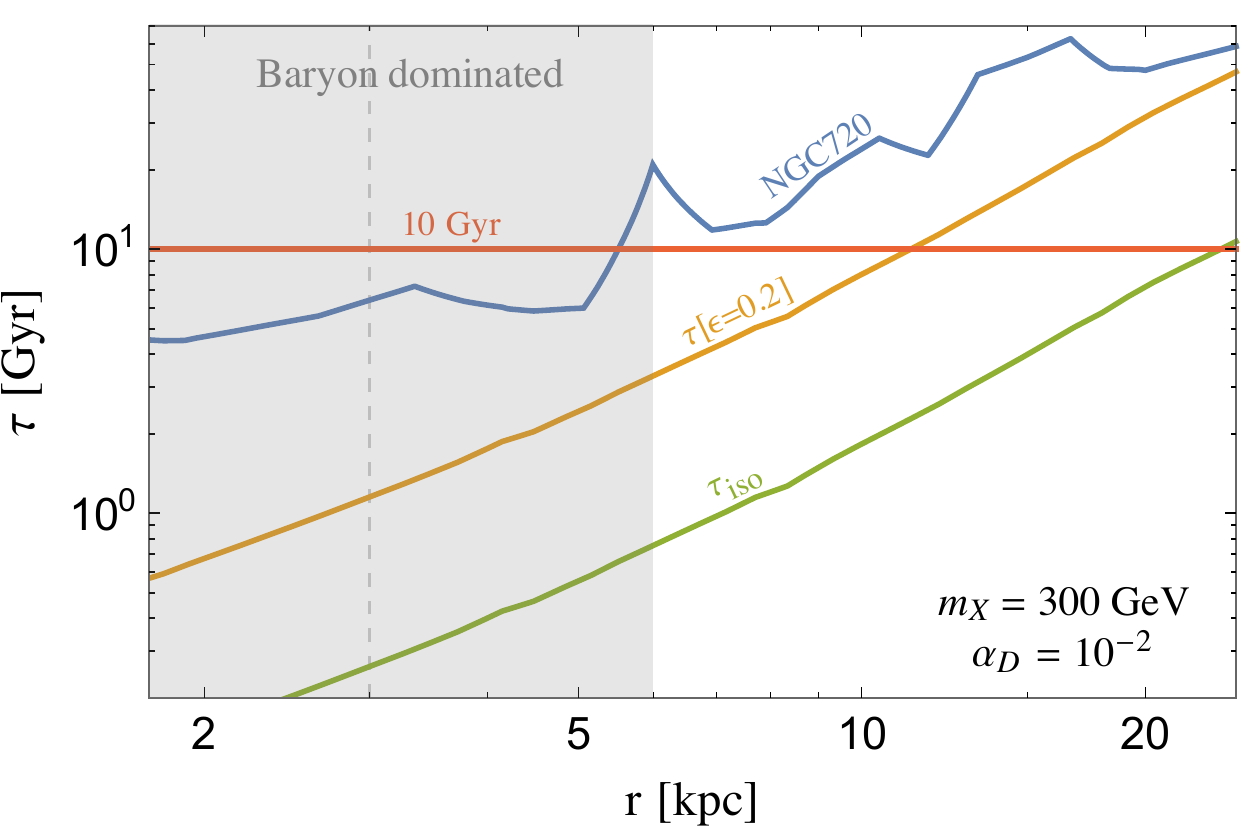}
  \end{center}
  \caption{\label{fig:ellsoph} Time to erase galactic ellipticity (caused by velocity anisotropy) through collisions of darkly-charged dark matter particles. The blue curve gives the time to reduce ellipticity in NGC720 down to $\epsilon(r)$ as given by the \emph{lower bound} from
figure~\ref{fig:NGCellipticity}. The orange curve shows the time it
takes to reduce ellipticity down to a \emph{fixed, average} $\epsilon =
0.2$ for each $r$. The green curve shows the timescale $\tau_{\rm iso}$ given by
eq.~(\ref{eq:taufeng}) (adopted from \citeref{Feng:2009mn}) where both
$\rho$ and $v_0$ are evaluated as functions of radius as given in \citeref{2011ApJ...729...53H}. We show the 10 billion year mark in red and the 3 kpc radius by a dashed vertical line. The
grey regions indicate radii for which baryons dominate the
gravitational potential of NGC720 and are thus not reliable for
constraining ellipticity of the dark matter component. Both panels use $m_X = 300$ GeV
but differ in their value of $\alpha_D$, with the left panel using $\alpha_D = 1.7\times 10^{-3}$ (which would
satisfy \citeref{Feng:2009mn}) and the right panel showing
$\alpha_D = 10^{-2}$ (which represents the bound from our calculation).}
\end{figure}

For example, the $2\sigma$ uncertainty region includes $\epsilon \sim
0$ for the inner most data point and therefore leads to virtually no
constraint at all for that radius. We study this effect in 
figure~\ref{fig:ellsoph}, where we
plot as a function of $r$ the time required to reach the ellipticity
corresponding to the
2$\sigma$ lower bound shown in figure~\ref{fig:NGCellipticity}.
We show our results for two
values of $\alpha_D$ with a sample mass $m_X = 300$ GeV.

\begin{figure}[h]
  \begin{center}
    \includegraphics[width=0.8\textwidth]{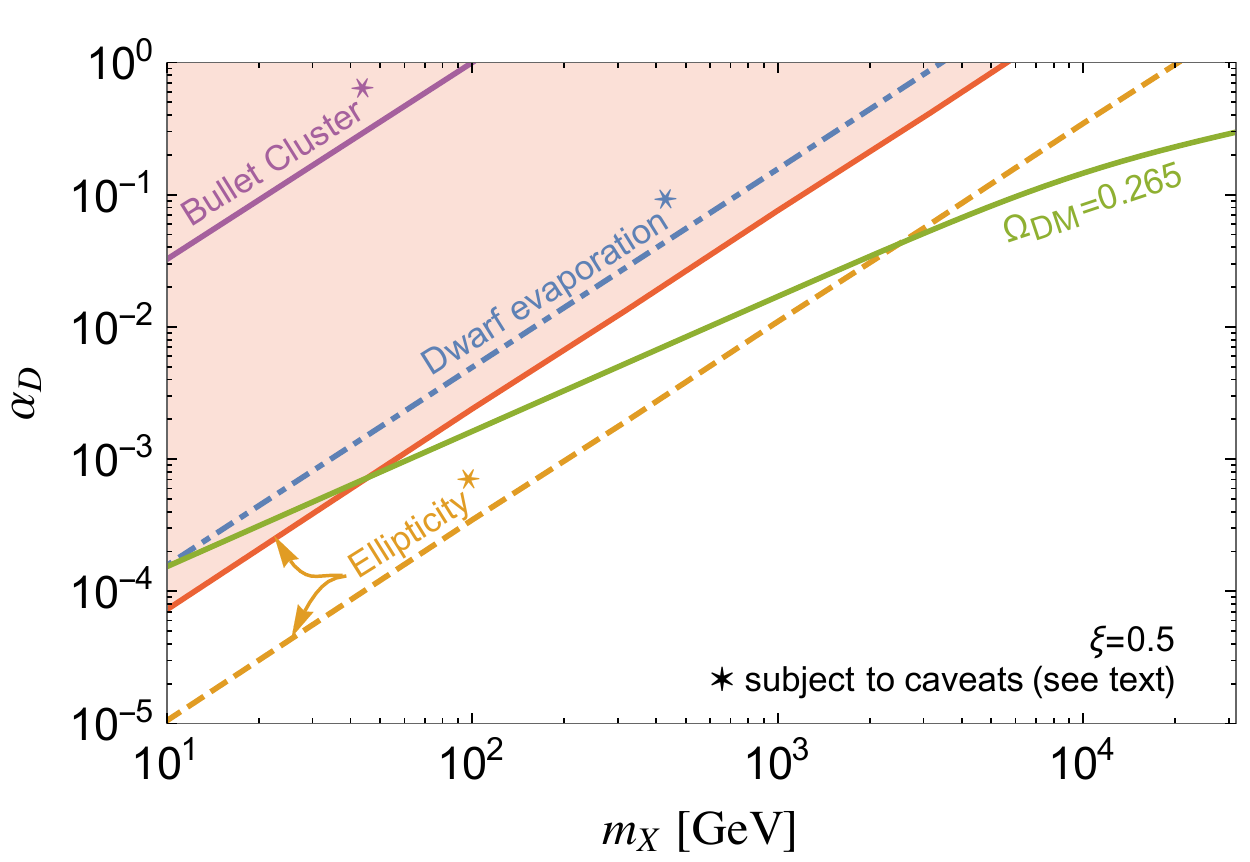}
  \end{center}
  \caption{\label{fig:constraints} Constraints on the darkly-charged
  dark matter parameter space in the $m_X-\alpha_D$ plane. Note that
  the constraints aside from relic abundance have the caveats
  discussed in the text and should not be taken as strict
  bounds on the parameter space. The ellipticity constraints
  (discussed in section~\ref{sec:ellipticity}) are presented as two
  curves: the original \citeref{Feng:2009mn} calculation [dashed
  yellow], and the more complete (though still uncertain) calculation
  that includes the radius dependent constraints on ellipticity from
  figure~\ref{fig:ellsoph} [red]. We also show the constraint from
  evaporation of Milky Way dwarf galaxies from
  \citeref{Kahlhoefer:2013dca} and discussed in
  section~\ref{sec:dwarf-survival} [dot-dashed blue].  We also display
  the Bullet cluster bound adopted from \citeref{Randall:2007ph} and
  discussed in section~\ref{sec:bullet-cluster} [purple]. Finally, we
  show the $m_X-\alpha_D$ curve for which the freeze-out mechanism
  discussed in section \ref{sec:relic_abund} produces the correct
relic density for darkly-charged dark matter [green], which includes
the effects of Sommerfeld enhancement.}
\end{figure}

Even when we include the effects of weakening bounds from
ellipticity at smaller radii we still obtain strong bounds from the
inner regions of the galaxy. However, these inner regions are
dominated by the mass of the baryonic component and so the ellipticity
of the local gravitational potential is dominated by the baryons. 
Furthermore, as found by \citeref{Kaplinghat:2013xca}, a thermalized self-interacting dark matter density profile is influenced
by the baryonic component. They find that the density ellipticity and the velocity anisotropy for dark matter
might not be correlated in the regions where baryons dominate the local potential.
As a result, we do not trust the ellipticity measurement as a
constraint on dark matter self-interaction for these inner radii.
\citeref{Buote:1993np} have constrained the mass
components of NGC720 as functions of radius. In their study the dark matter component
becomes dominant in mass only after $r \sim 6$ kpc. As a result we
will cutoff the constraints on $\alpha_D-m_X$ from the ellipticity
measurement at this radius $r  = 6$ kpc. Figure~\ref{fig:constraints} shows the
resulting ellipticity constraint in the $\alpha_D-m_X$ plane together with other constraints we discuss in the following sections. We note that even this revised constraint is subject to uncertainties, as mentioned in section \ref{sec:ellipticity}.

\subsection{Dwarf Galaxy Survival}
\label{sec:dwarf-survival}

We now turn  to other potential constraints. The strongest such constraint is from dwarf galaxy
survival as they orbit in the halo of the host galaxy. With too
strong an interaction, dwarf galaxies will be stripped as they pass
through a halo.  \citeref{Kahlhoefer:2013dca} derived a reasonably strong constraint on dark
matter--not quite as strong as the purported ellipticity constraint
but stronger than that from the Bullet Cluster, for example. Again,
they found that numerous soft scatterings dominated over a single hard
scattering. As presented in \citeref{Kahlhoefer:2013dca}, the constraint reads 
\begin{equation}
{\alpha_D^2 \over m_X^3}<10^{-11} \, {\rm GeV}^{-3}.
\label{eq:kahlhoefer}
\end{equation}

We believe this constraint is also somewhat
overstated, both for technical reasons that reduce the bound slightly
and for other reasons that could be very interesting, but require a
more careful analysis that we delay to a future publication.
We do not present the full calculation but rely on the calculation for
ellipticity as presented above. The technical disagreements are as
follows:

\begin{itemize}
\item The same consideration about the cutoff in the infrared
  logarithm applies here. Putting in the inter-particle spacing rather
  than the Debye wavelength weakens the numerical value of the bound in eq.~(\ref{eq:kahlhoefer}) by about a factor of 2/3.

\item \citeref{Kahlhoefer:2013dca} integrates $\cos \theta_\text{cm}$
  on the interval $[0,1)$ and then multiply by a factor of 2. This overestimates the cross-section by a factor of 2.

\item The analysis assumed a dark matter density of  $10^{-26} {{\rm g} \over {\rm cm}^3}$. 
Though approximately correct, this is probably a slight overestimate of the density in the region for the relevant dwarf galaxies. Putting
in the Carina dwarf galaxy we found the density overestimated by a factor of 3, assuming the dwarf does not spend much time at smaller radius, which was assumed for a conservative bound.
\end{itemize}
Putting all these factors together:
\begin{equation}
f = \underbrace{\frac{3}{2}}_{\log\Lambda} \times \underbrace{2}_{d\sigma/d\Omega} \times \underbrace{3}_{\rho} = 9
\end{equation}
introduces additional uncertainty -- up to an order of magnitude in the cross section
and a factor of a few in the mass, for example. However,
\citeref{Kahlhoefer:2013dca} have chosen to understate their bound by
a factor of about 4, and so numerically the bound does not change by
more than a factor of 2.

However, the bound is in fact even more subtle. The above calculation
takes into account multiple scattering of an individual dark matter
particle with multiple halo particles. But it neglects the
interactions of the dark matter particles inside the dwarf -- where
dark
matter is far denser and slower -- leading to core formation and
potential core collapse, as discussed in section~\ref{sec:superstrong}
below.
Allowing for this effect redistributes
energy so that rather than eventually lifting a particle to escape
velocity, the multiple scatterings of all dark matter particles can
redistribute dark matter in the dwarf galaxy itself. How this does so
requires a full detailed calculation. But it seems likely that dwarfs
will puff out so that smaller dwarfs will become bigger in size -- and also
slow down core collapse to make them less dense in their cores. This
is clearly of interest to evaluate further. For the purposes of this paper, we note only that  these considerations may weaken the bound on darkly-charged dark matter.

\subsection{Merging Cluster Constraints}
 \label{sec:bullet-cluster}

Merging galaxy clusters are unique probes of the large velocity limit of the dark matter self-interaction cross section \cite{Markevitch:2003at,Randall:2007ph,Dawson:2012fx,Mohammed:2014iya,Dawson:2014jca,Kahlhoefer:2013dca,Kahlhoefer:2015vua,Harvey:2015hha,Massey:2015dkw,Randall:2016nxa,Robertson:2016xjh}. Detailed observations of merging clusters lead to different types of dark matter constraints, including those based on the gas and dark matter offset, the high velocity of the subclusters after the initial collision, the survival of distinct dark matter-dominated mass peaks after the collision, and the possible presence of an offset between the mass peak and the galaxy centroid. For the Bullet cluster \cite{Markevitch:2003at,Randall:2007ph}, the two latter techniques yield the most stringent constraint. This result was used in \citeref{Feng:2009mn} to impose an additional bound on darkly-charged dark matter, albeit a very weak one due to the large relative velocities of the colliding clusters. We include this constraint in figure \ref{fig:constraints}. However, this much weaker bound is itself subject to mitigating factors.  For instance, we know only the dark matter and baryonic densities after the merger and the initial conditions are entirely unknown.  It is not even known whether the ratio of baryonic to dark matter components is the same for the two merging clusters. For this reason, we don't know enough about the fraction of the initial dark matter component that can be removed through scattering. Recently, Ref.~\cite{Robertson:2016xjh} has shown that uncertainties associated with the measurement technique used to extract the dark matter-galaxy offset can even supersede the above concerns, leading to a weakening of this already feeble bound. 

As more merging clusters are analyzed \cite{Dawson:2012fx,Mohammed:2014iya,Dawson:2014jca,Kahlhoefer:2013dca,Kahlhoefer:2015vua,Harvey:2015hha,Massey:2015dkw,Randall:2016nxa} and systematics become better understood, they could eventually provide improved constraints on the dark matter self-interaction strength. Still, given the strong suppression at large velocities of the transfer cross section for darkly-charged dark matter, it is unlikely that merging clusters will ever provide the strongest bound on our model.

\section{For the Future}
\label{sec:future}

We have presented an updated calculation of the rate at which one
population of stars would transfer velocity to another, or for the
case of interest the rate at which large components of velocity would
transfer energy to other velocity components. We have shown that the
constraint on the interaction cross section is weaker than previously
claimed, allowing for lighter dark matter particle or a larger charge --
assuming a massless photon in the dark sector.  In this section, we
present a number of open issues that would be essential to pinning down
the strongest constraint on darkly charged dark matter and perhaps
indicating a way forward in terms of the search for interactivity.

As explained, the ellipticity constraint as calculated might still be overly
strong. We know that the connection between an isotropic distribution of velocities and an isotropic potential is robust only for isolated and relaxed halos. The constraint we presented was based on the assumption that the timescale for isotropizing velocities is similar to that necessary to remove any anisotropies in the gravitational potential itself. A dedicated cosmological simulation of a dark matter halo with Coulomb-like interactions would probe the interplay between velocity and potential isotropy, and assess whether our assumption is justified. A potentially interesting direction will be to see if
radial dependence can help distinguish the two sources of anisotropy, at least on a
statistical basis. Ellipticity changes with radius, and it would be
less likely for strong changes to be associated with anisotropy in
velocities. 

Another extremely interesting area for study
is the interaction of the dwarf galaxies with their host haloes. Dark
matter self-interaction can inject heat into the dwarves, and
can interplay with self-interaction within the dwarf (we consider the
strong self-interaction within dwarves in more detail in
section~\ref{sec:superstrong}).
This will lead to modifications of the mass function and internal structure of dwarf galaxies 
should there be significant interaction, affecting the predictions that should be compared with observations.  For instance, we expect to find fewer dwarf
galaxies than would otherwise predicted since some will evaporate.
Furthermore we expect the density profile of dwarf satellite galaxies
to evolve toward bigger, less dense objects with potential
implications for core-cusp issues \cite{2010AdAst2010E...5D,Gentile:2004tb,Walker:2011zu} and the Too Big to Fail problem \cite{BoylanKolchin:2011de}.
Again, we leave this for further work. 

The third type of constraint we discussed arises from merging cluster
observations. Due to the large typical velocities of these merging
systems, they yield weaker bounds on darkly-charged dark matter than
the dwarf galaxy survival constraint. As outlined above, modeling
uncertainties and possible systematics in the measurement techniques
can further weaken the constraints, although we can expect those to
come under better control as more clusters are analyzed. It is of
interest to see what we can learn from these types of objects in the
future as well, particularly if there are separate dark matter
populations with various degrees of interactivity. As emphasized in
Ref.~\cite{Kahlhoefer:2015vua}, it might even be possible to
distinguish dark matter models with a contact-type interaction from
models displaying long-range forces such as darkly-charged dark
matter, an intriguing possibility. 

One of the most interesting aspects of darkly-charged dark matter, and one that is the hardest to quantify, is the possible presence of collective plasma effects within the dark sector. The massless dark photon allows for a range of phenomena such as plasma waves, magnetic instabilities, and shocks \cite{2015PhLB..749..236H,Sepp:2016tfs} that all could be relevant to small-scale structures. For instance, Ref.~\cite{Ackerman:2008gi} determined that the Weibel plasma instability could be relevant for the entire darkly-charged dark matter parameter space. Detailed magnetohydrodynamics simulations would be required to fully understand whether this can lead to new constraints on this darkly-charged dark matter.

As we have already mentioned, many of the arguments we present rely on
assumptions about the unknown initial state of a given galaxy or
clusters of galaxies.  In the case of a galaxy, we have only one well
measured system -- NGC720.  However, having a larger statistical sample of
these objects would help make the constraints more robust 
and, to a certain extent, eliminate the initial state dependence. 
This is especially the case for elliptical galaxies. Measurement of
the ellipticity of dwarf galaxies can potentially yield a stronger
constraint since we expect our self-interaction cross section to be
bigger in colder systems.

\subsection{Cores in Very Strongly Interacting Systems}
\label{sec:superstrong}

One final consideration applies only in the presence of interactions
as strong as those we have shown to be permissible.
Self-interacting systems can affect the structure of dark matter
haloes at cluster, galaxy, and dwarf galaxy scales. In principle,
self-interacting systems can lead to
core formation by allowing heat to flow from the outer part of a
galaxy to the center and thereby expanding a central, initially cuspy
region. With stronger interactions still, core collapse can occur on
galactic time scales \cite{Balberg:2002ue,Koda:2011yb}.

This type of gravothermal collapse arises in a number of
astrophysical settings such as globular clusters, and during star and planet formation. The core collapse in dark matter haloes can be
studied in analogy with these systems.
The rate of collapse is determined by the rate of
heat transfer from the center of the system to outside, which depends
on the temperature and density of the system, which can be very
different in different cases. 
In some cases nuclear forces provide an additional heat source.

For dark matter haloes the rate of heat loss can be offset by 
the presence of a host halo or from cosmological infall, which can
provide an external source of heat input 
slowing down or stopping
gravothermal collapse~\cite{Ahn:2004xt}. This can lead to a difference
in halo properties for dwarves in host galaxies and isolated dwarves.

A critical
quantity characterizing the thermal evolution of the system is the
Knudsen number
\begin{align}
  Kn
  &=
  \frac{\lambda}{R}
\end{align}
defined as the ratio of mean free path $\lambda$ to the size of the object $R$. This
tells us how effective scattering is within an object of a given size. For small cross sections and hence large Knudsen numbers, $Kn > 1$,
the conductivity is inversely proportional to $Kn$. Specifically, $Kn \gg 1$ ($\sigma \to 0$) corresponds to essentially 
collisionless dark matter. For $Kn \gtrsim 1$, heat conduction is
effective and leads to core formation 
on galactic timescales through transfer of heat from the outer parts
of the halo to the inside. For even stronger cross sections,
the direction of the heat flow can reverse \cite{Balberg:2002ue}
and the core can undergo a collapse exhibiting gravothermal
catastrophe.  But for $Kn\ll1$, (small mean free path),  heat
conduction is suppressed, and hence both core formation and core
collapse are inhibited. Therefore, for extremely strong interactions the
system might more closely resemble the non-interacting 
one~\cite{2003JKAS...36...89A,Ahn:2004xt}.

This leads to the rather remarkable possibility that sufficiently
strong interactions can lead to density profiles different from either
a non-interacting system or the SIDM models that have already been
studied.  In our case, the cross section of interest is highly
velocity dependent. In various systems, dwarf galaxies, galaxies and clusters, we calculate a cross section for
self-interactions of order

\begin{align}\label{eq:cross_section_regime}
  \frac{\sigma_T}{m_X} 
  &= 
  \frac{8\pi \alpha_D^2}{m_X^3 v^4}\log\Lambda
  =
  \left\{
  \begin{array}{ll}
  1.7\times10^{4}\;\; \,
  \frac{\text{cm}^2}{\text{g}} 
  \left(\frac{\alpha_D}{2.5\times10^{-3}}\right)^2
  \left(\frac{100\;\text{GeV}}{m_X}\right)^3
  \left(\frac{\log \Lambda}{45}\right) 
  \left(\frac{30\;\text{km/s}}{v}\right)^4  
  & \text{Dwarf galaxies}\\
  2.1\times 10^{0}\;\; \,
  \frac{\text{cm}^2}{\text{g}} 
  \left(\frac{\alpha_D}{2.5\times10^{-3}}\right)^2
  \left(\frac{100\;\text{GeV}}{m_X}\right)^3
  \left(\frac{\log \Lambda}{60}\right)
  \left(\frac{300\;\text{km/s}}{v}\right)^4  
  & \text{Galaxies}\\
  2.0\times10^{-2} \,
  \frac{\text{cm}^2}{\text{g}} 
  \left(\frac{\alpha_D}{2.5\times10^{-3}}\right)^2
  \left(\frac{100\;\text{GeV}}{m_X}\right)^3
  \left(\frac{\log \Lambda}{72}\right)
  \left(\frac{1000\;\text{km/s}}{v}\right)^4  
  & \text{Clusters.}
  \end{array}
  \right.
\end{align}

The interaction cross section in dwarf galaxies is several orders of
magnitude greater than the value for which Ref.~\cite{Vogelsberger:2012ku}
found evidence for core collapse.  For these values of the parameters,
we can estimate the Knudsen numbers in various systems,
\begin{align}
  Kn
  &\simeq
 \left\{
 \begin{array}{ll}
  10^{-3}  
  \left(\frac{1\,\text{kpc}}{R}\right)
  \left(\frac{9\,\text{GeV/cm}^3}{\rho}\right)
  \left(\frac{1.7\times 10^{4} \text{cm}^2/\text{g}}{\sigma_T/m_X}\right)
& \text{Dwarf galaxies}\\
  10^{1}\;\; 
  \left(\frac{30\,\text{kpc}}{R}\right) 
  \left(\frac{0.3\,\text{GeV/cm}^3}{\rho}\right)
  \left(\frac{2.1 \text{cm}^2/\text{g}}{\sigma_T/m_X}\right)
  & \text{Galaxies}\\
  10^{5}\;\; 
  \left(\frac{10\,\text{Mpc}}{R}\right) 
  \left(\frac{9\times 10^{-6}\,\text{GeV/cm}^3}{\rho}\right) 
  \left(\frac{2.0\times 10^{-2} \text{cm}^2/\text{g}}{\sigma_T/m_X}\right)
  & \text{Clusters.}
 \end{array}
\right.
\end{align}

We see that for dark matter as light and as strongly interacting as we
have found is allowed, we can be in the very small $Kn$ regime for
dwarf galaxies, with a transition to the more standard SIDM scenario
as velocity increases and density decreases. It is worth noting
that if the constraints on $\alpha_D$ were an order of magnitude
stronger, the smallest Knudsen numbers in the dwarf galaxies would be
$\mathcal{O}(1)$. Therefore, the weaker bounds in $\alpha_D$ that we
have found open up the small Knudsen number region in dwarf galaxies.
It is unclear without a more
detailed analysis what the consequences will be, leaving open the
interesting possibility that this velocity-dependent cross section can
evade other bounds but have interesting consequences in dwarf
galaxies. 

One way to interpret this result is in terms of a ``cut-off'' beyond
which the system goes over to an effective more weakly interacting
theory, presumably by coarse-graining over the mean-free path. Because
a cutoff is automatically imposed by the strong interactions that
occur at small velocity, this opens the possibility of fitting to
observed galaxy and galaxy cluster shapes over a wide range of scales.
Ref. \cite{Kaplinghat:2015aga} worked with available data to fit cores
to different-sized objects, ranging from galaxy clusters to dwarf
galaxies.   
Our cross section for clusters ($0.02$ cm$^2$/g) is
somewhat smaller than their best fit ($\sim0.1$ cm$^2$/g), but
may be consistent with core formation in clusters. The cross section
at galaxy scales is on the larger side (1--10 cm$^2$/g), but again may
be consistent within uncertainties. 
However, our cross section at the low velocity scales in dwarf
galaxies
is huge ($10^4$ cm$^2$/g), since it scales as $1/v^4$.
In order to fit to a much lower cross section (1 cm$^2$/g),
Ref.~\cite{Kaplinghat:2015aga} imposed a mass for the mediator
to cut off the cross section at low velocities.
It is interesting to note that there may be no need for such a mass
for the mediator since the short-mean-free path serves as a dynamic
cutoff at small velocity. In fact, refs.~\cite{2003JKAS...36...89A,Ahn:2004xt} seems to indicate an approximate duality $ Kn\to 1/Kn$ between the strongly and weakly interacting regimes. It will be interesting to explore the robustness of this rough symmetry. Furthermore, the authors of Ref.~\cite{2003JKAS...36...89A} indicated that the value of $10^4$ cm$^2/$g that are relevant for the dwarf galaxies in our model (see eq.~\eqref{eq:cross_section_regime} above) might lead to core formation similar to that of standard SIDM. 

Here, we have explored just a few target systems and velocities. In reality, there is a broad range of objects with velocity dispersions spanning the values between those of the large galaxies and dwarf galaxies that we listed in eq.~\eqref{eq:cross_section_regime}. In principle, our model makes a strong prediction for the properties of these intermediate objects since the cross section increases dramatically as we go from large galaxies to the smallest dwarves. Over a significant range, the dark matter interaction strength will be sufficiently weak that we expect results can be found reliably at this point and compared to data. The strong velocity dependence of a cross-section provided by a massless mediator should allow for the most stringent tests of this model, and perhaps ways to even discover the massless dark photon processes in the future. 

Clearly the photon-mediated velocity-dependent cross section provides
an extremely rich interesting system. It is remarkable that the necessary
cutoff appears to be automatically imposed by the strong interactions that occur
at small velocity. In a future analysis we envision imposing an
effective theory, in which strongly interacting regimes would be
replaced by more weakly interacting ones by coarse-graining over mean free
paths.  This should provide an approximate realization of this system
and might even allow for a fit to cores over a wide range of scales.

For now, we note that the core constraint  is
an important one.  It nonetheless does not currently rule out the
interesting darkly-charged dark matter scenario that we envision.

\section{Conclusions}
\label{sec:conclusions}
We have argued that darkly-charged dark matter, where dark matter
experiences a long range force not experienced 
by ordinary Standard Model matter,  is a viable possibility with
extremely rich phenomenology.  We have shown that the allowed
parameter space is considerably less restrictive than previously
assumed and dark matter can be as light as the weak scale, $100$ GeV,
and still experience significant interactions: $\alpha_D = 2.5\times
10^{-3}$, where even this constraint, which comes from relic abundance, can
change in more elaborate models. The renewed parameter space is important as it says
that dark matter, which we take to be relatively inert, can
conceivably have reasonably strong interactions and have mass
comparable to that of known Standard Model particles and still be
consistent with known observations.  Intriguingly, the weaker bounds
also open up parameter space for novel dark matter halo dynamics such
that dark matter in dwarf galaxies can be strongly self-interacting.
Such interactions inhibit heat flow over scales larger than the mean
free path, introducing a dynamical cutoff to the self-interactions.  

In a companion paper \cite{Agrawal:2017rvu}, we consider the case where
in addition to the weak-scale $X$ dark matter particle, there is a
light dissipative component which can lead to a dark matter disk.
Unlike the previous study~\cite{Fan:2013tia,Fan:2013yva} where the
halo comprised of a separate CDM species, all of dark matter would be
composed of charged components. This has additional interesting
consequences for the formation of structure in the early universe.

Although current constraints are relatively weak, they lie at the
boundary of the favored region where darkly-charged dark matter can be
a thermal relic. Consequently, future observations may be able probe
this extremely promising region and provide the opportunity to learn
more about the nature of dark matter. In particular, charged dark
matter can affect the distribution of structure in the Universe and
might ultimately provide a better match to data.  Given our lack
of knowledge about the nature of dark matter, darkly-charged dark
matter is a simple possibility worth considering which we might have
experimental access to in the near future.

\begin{acknowledgments}
  We thank Sasha Brownsberger, Doug Finkbeiner,
  Gil Holder, Annika Peter, Subir Sarkar, Neal Weiner, Linda Xu, Hai-Bo Yu, and Kathryn Zurek for useful discussions. We thank Manoj Kaplinghat for useful comments on an earlier version of this manuscript. F.-Y. C.-R.~acknowledges the support of the National Aeronautical and Space Administration ATP grant NNX16AI12G at Harvard University. This work is supported by NSF grants PHY-0855591 and PHY-1216270. We would like to thank the Aspen Center for Physics, the Mainz
  Institute for Theoretical Physics, and David Rubenstein for hospitality during the completion of this work.
\end{acknowledgments}

\appendix

\section{Scattering Cross-section}
\label{sec:xsection}
The process for $XX \to XX$ and $\bar{X}\bar{X} \to \bar{X}\bar{X}$
corresponds to M\o ller scattering. The process responsible for
$X\bar{X} \to X\bar{X}$ represents Bhabha scattering. The cross
section, in the CM frame, for M\o ller process is: 
\begin{align}
  \frac{d\sigma}{d\Omega} 
  &=
  \frac{\alpha_D^2}{2 s}
  \left(
  \frac{s^2 + u^2 +8m^2t -8m^4 }{t^2}
  +\frac{s^2 + t^2 +8m^2u -8m^4 }{u^2}
  +\frac{2s^2 -16m^2 s +24m^4}{ut}
  \right).
\end{align}
where $s,t,u$ are the usual Mandelstam variables. They can be
parametrized as,
\begin{align}
  s &= 4m^2 + 4m^2 v_\text{cm}^2 + \mathcal{O}(v_\text{cm}^4)\\
  t &= -\frac{1}{2}(s-4m^2)(1-\cos \theta_{\text{cm}}) \\
  u &= - \frac{1}{2}(s-4m^2)(1+\cos \theta_{\text{cm}})
\end{align}
The Bhabha scattering cross-section can be obtained by $s
\leftrightarrow u$ crossing symmetry. To the lowest order in
$v_{\text{cm}}$ we recover the differential
cross section for M\o ller process:
\begin{equation}
 \frac{d\sigma_{\text{cm}}}{d\Omega}=\frac{\alpha_D^2 }{4 m^2 v_{\text{cm}}^4 (1-\cos \theta_\text{cm})^2} \frac{(1+3\cos^2 \theta_{\text{cm}})}{(1+\cos \theta_\text{cm})^2}
\end{equation}
Notice that this cross-section diverges for both $\theta_\text{cm} =
0$ and $\pi$. However, because of the indistinguishable final state
particles, we only consider the range $0 < \theta_\text{cm} <\pi/2$.
Similarly the lowest order in $v_{\text{cm}}$ the Bhabha process
gives:
\begin{equation}
\frac{d\sigma_{\text{cm}}}{d\Omega}=\frac{\alpha_D^2 }{4 m^2 v_{\text{cm}}^4 (1-\cos \theta_\text{cm})^2}
\end{equation}
This process has only singularity for $\theta = 0$ and the full range
$0<\theta_\text{cm} < \pi$, because $X$ and $\bar{X}$ are
distinguishable. Since our cross-sections will be dominated by their
singular behavior at $\theta_\text{cm} = 0$, we only need to use the
$\theta_\text{cm}^{-4}$ terms for which the M\o ller and Bhabha
processes agree:
\begin{align}
\frac{d\sigma_{\text{cm}}}{d\Omega} &\sim \frac{4\alpha_D^2}{m^2 |v_1-v_2|^4 (1-\cos \theta_\text{cm})^2} 
=\frac{\alpha_D^2}{16m^2 v_\text{cm}^4 \sin^4 \frac{\theta_\text{cm}}{2}}
\end{align}
As a side note, we remark that this cross section agrees with that used in \citeref{Kahlhoefer:2013dca} in the $\theta \to 0$ limit.

We note that the cross sections above are obtained in the Born
approximation; a more appropriate 
regime for our calculation would be the classical
regime~\cite{Tulin:2012wi,Tulin:2013teo}, where multi-photon processes
also contribute. For the massless mediator, however, the Born and
classical differential cross sections are the same, and the difference
only appears in the IR cutoff used to calculate the momentum
transfer cross section.
For a massive mediator, the mass provides an automatic IR cutoff so
that the classical and Born calculations 
differ~\cite{Tulin:2012wi,Tulin:2013teo}.
Our computations in the
text agree with the classical cross sections
in Ref.~\cite{Khrapak:2003kjw} (except we use the inter-particle spacing as
the screening length).

\section{Differential equation for velocity anisotropy}
\label{sec:odeapp}
The goal of this Appendix is to simplify the expression:
\begin{align}
   -\frac{3 \dot{v}_c}{ v_c^4}= &
  \frac{32 \pi^3\sqrt{\pi}\alpha_D^2}{m^2 n_c }
  \int    \frac{ d v_2}{ v_2} \frac{dx}{(1-x)^2}
  \left[f(0) f(v_2) - f(v_2\sqrt{(1-x)/2}) f(v_2\sqrt{(1+x)/2})\right]
  \label{eq:eqF}
\end{align}
Let us first consider the term in the brackets:
\begin{align}
  G=& f(0) f(v_2) - f(v_2\sqrt{(1-x)/2}) f(v_2\sqrt{(1+x)/2})\notag\\\notag
    =& \left[\frac{n_c}{v_c^3}+ \frac{n_h}{ v_h^3}  \right]\left[\frac{n_c}{v_c^3} e^{-\frac{v_2^2}{v_c^2}} + \frac{n_h}{v_h^3} e^{-\frac{v_2^2}{v_h^2}} \right]-\left[\frac{n_c}{v_c^3} e^{-\frac{v_2^2(1-x)}{2v_c^2}} + \frac{n_h}{v_h^3} e^{-\frac{v_2^2(1-x)}{2v_h^2}}\right]\left[\frac{n_c}{v_c^3} e^{-\frac{v_2^2(1+x)}{2v_c^2}} + \frac{n_h}{v_h^3} e^{-\frac{v_2^2(1+x)}{2v_h^2}}\right]
\end{align}
All the terms proportional to $n_c^2$ and $n_h^2$ cancel because each Gaussian is already an equilibrium distribution. Therefore only cross-terms are left:
\begin{equation}
G = \frac{n_h n_c}{ v_h^3 v_c^3}\left( e^{-\frac{v_2^2}{v_c^2}} +e^{-\frac{v_2^2}{v_h^2}}- e^{-\frac{v_2^2(1+x)}{2v_c^2}}e^{-\frac{v_2^2(1-x)}{2v_h^2}} - e^{-\frac{v_2^2(1-x)}{2v_c^2}} e^{-\frac{v_2^2(1+x)}{2v_h^2}}\right)
\end{equation}
Since we are interested in the $\theta' \to 0$, i.e. $x\to1 $ limit, we expand $G$ around $x=1$:
\begin{equation}
G = n_h n_c \frac{v_2^2}{2} \frac{(v_c^2- v_h^2)}{v_c^5 v_h^5} \left( e^{ - v_2^2/v_c^2} - e^{ - v_2^2/v_h^2}\right)(1-x) + \mathcal{O}\left[(1-x)^2\right].
\label{eq:eqG}
\end{equation}
Combining equations~(\ref{eq:eqF})~and~(\ref{eq:eqG}) we obtain
\begin{align}
 \dot{v}_c= &
  -\frac{16 \pi^3\sqrt{\pi}\alpha_D^2 n_h }{3m^2 }\frac{(v_c^2- v_h^2)}{v_c v_h^5}
  \int     v_2 d v_2 \left( e^{ - v_2^2/v_c^2} - e^{ - v_2^2/v_h^2}\right) \int  \frac{dx}{(1-x)}
\end{align}
Finally, the integral over $v_2$ is simple and we obtain a simple expression for $\dot{v}_c$:
\begin{align}
 \dot{v}_c= &
  -\frac{8 \pi^3\sqrt{\pi}\alpha_D^2 n_h }{3m^2 }\frac{(v_c^2-
  v_h^2)^2}{v_c v_h^5} \int_0^{\cos^{-1}\theta_\text{min}}  \frac{dx}{(1-x)}
\end{align}

\bibliographystyle{JHEP}
\bibliography{ref.bib}

\end{document}